# Mobile crowdsourcing - activation of smartphones users to elicit specialized knowledge through worker profile match

Oskar Jarczyk[1]

[1] Polish-Japanese Institute of Information Technology, Warsaw
oskar.jarczyk@pjwstk.edu.pl

**Abstract**

Crowdsourcing models applied to work on mobile devices continuously reach new ways of solving sophisticated problems, now with a use of portable advanced devices, where users are not limited to a stationary use. There exists an open problem of quality in crowdsourcing models due the inexperienced or malicious workers. In this paper, we propose a model and a short specification of a platform for a bundled widely available crowdsourcing mechanism, which tries to utilize workers individual characteristics to maximum. Analyzed solution relies on geographical data classified by localization category. Secondly, we profile mobile workers by precisely analyzing their activity history. Results of this research will make an impact on better understanding the latent potential of mobile devices users. It makes for not only better quality in results, but also opens a possibility of implementing a "twitch crowdsourcing" or emergency relief systems. Special tasks assigned to owners of mobile devices can help those, which are in need of help, making them the task creators.

## 1 Introduction

Crowdsourcing can be defined as the outsourcing of simple small tasks traditionally performed by specific individuals to an undefined large group of people or a community (crowd) through an open or targeted call for collaboration. It is not a new model and it have some predecessor in the history. The very beginning was a quiz by Sir Francis Galton, when he asked the crowds about possible weight of the ox, nobody knew the exact answer, but the mean value of replies was very close to the answer. There are different types of crowdsourcing, including: wisdom of the crowd, crowdfunding, microwork, macrowork, crowdvoting, creative crowdsourcing and inducement prize contests. Additionally, it is also possible to recognize crowdsearching and implicit crowdsourcing. In the era of mobile devices (i.e. smartphones, tablets, smart devices) it is possible to reach wide audience of people being both in the move and stationary at one point of time, being able to respond to job almost immediately, on a device characterized with a geological location and additional sensors. Moreover, mobile users can be creators of crowdsourcing tasks as well, which makes it a fast robust bi-directional network. Crowdsourcing has been a successful model of gathering knowledge of crowds already for many years. What differs mobile crowdsourcing from the 'standard' crowdsourcing model are some optional assumptions. Worker can be perceived *a prori* as a person being currently in a move. He or she is providing more data from the environment through the device sensors, and finally - messages and task assignment are delivered instantly.

## 1.1 Practical appliance

Numbers justify necessity of focusing on this market. 82 million people in the US have a tablet. More than 41.1% (26 million) of people in the UK will use a tablet regularly in 2014. By end of 2014, there will be more mobile-connected devices on Earth than people. Those numbers explain clearly the reason for an increase of companies dealing with crowdsourcing implemented to work with mobile devices. Twitter is told to be a father of basic mobile crowdsourcing, but first specialized systems where created somewhere after 2011, in 2013 they become quite popular. For better understanding this research specification, understand the definition of a crowdsourcing task and a worker.

# 2 Related work

Most recent and similar research was held by Vaish, Wyngarden et.al. and presented in the article "*Twitch crowdsourcing: crowd contributions in short bursts of time*" (2014). Researchers present twitch crowdsourcing - a crowdsourcing model with quick contributions, that can be completed in one or two seconds. Person is asked e.g. for authoring a census of local human activity, rating stock photos, or extracting structured data from Wikipedia pages. Jobs vary, yet most of the tasks are basic enough to close them in 2 seconds, which a help of efficient and simple user interface, which shows the most important data immediately. The user median activity took *1.6* seconds, incurring no statistically distinguishable costs to unlock speed or cognitive load compared to a standard slide-to-unlock interface. There are also couple of more papers which are worth mentioning, for better understanding of possible solutions. In a paper titled "*Crowdsourcing in Mobile: A Three Stage Context Based Process*" Afridi (2011) explains general ideas of implementing crowdsourcing on mobile systems.8 Author research deals with the issues of uncertainty and quality of social mobile computing systems. It suggests the relationship between context and crowd sourcing activity. This understanding can enhance the future and existing applications in a way that it can optimize systems response, adapt the components and decrease the uncertainty. In paper "*iMac: strategy-proof incentive mechanism for mobile crowdsourcing*" Feng et.al. (2013) propose a strategy-proof incentive mechanism based on Vickrey-Clarke-Groves (VCG) mechanism. Article "*A crowdsourcing based mobile image translation and knowledge sharing service*", Liu et.al. (2010), shows discussion how useless answers – due misunderstandings – influence results, and why proofreading may solve this problem. In the work "*Designing novel mobile systems by exploiting sensing, user context, and crowdsourcing*", Yan et.al. (2012) author explains the term of sensing-driven mobile systems. Scholar explains that smartphones are used to sense personal context information. Secondly, smartphones are exploited to collect sensing data of the physical world and enable applications including traffic monitoring, environmental monitoring, and others. He propose a decision engine to improve the responsiveness of mobile OS by 2x times. Next, scientist improves image search accuracy on centralized servers.

## 2.1 Research on impact of adding geolocation

"Context-Aware Mobile Crowdsourcing", Tamilin et.al. (2012) presents a system for bi-modal interaction channel between citizens and public administration. Researchers define a mobile crowdsourcing task as a tuple of a task context scoping its applicability, and an action the task consists of. Scholars defined a distance between a user context and a task context. Finally, they made an algorithm to optimally present tasks to users and preserve mobile devices resources.

# 3  Problem description

Supporting mobile crowdsourcing models require to create a good background propagation of information. It is characterized by: *availability of clients* (Av), *transfer usage* (Tu), *failure rate* (FailR), making them the benchmarks of the data transportation. Main goals are yet directly connected with the crowdsourcing implementation – worker effectiveness and proper task completion. Let's define worker location as a pair of two elements. First is geolocation as a point in 2D geographical location. Second is a classification of a location from a given set (e.g. work place, shopping mall, train). Another aspect included in the list of dimensions is a worker profile. It is calculated from the history of his/her activity. We want to use worker location and worker profile to predict best possible match for all future crowdsourcing tasks. We benchmark this through the quality model described in paragraph 4.2. Below are hypothesis that research will confirm or reject.

**Hypothesis 1**: Mobile crowdsourcing methods are an effective way of coping with local emergency situation.

Lot of previous research confirmed successful usage of social media in relief of emergency situations. There was also numerous research on helping an individual with copying with personal or group problems, as well on supporting good relations between groups of interest.

**Hypothesis 2**: Profiling mobile crowdsourcing workers through their activity and common geolocation will improve both worker quality and worker response time.

Different workers have different crowdsourcing characteristics (response time, correctness) basing on their current "busyness". Person waiting for a train will be more likely to participate in task resolution then an university student during finals.

**Hypothesis 3**: Adding geolocation attribute will make for higher quality in mobile crowdsourcing models.

Intuition suggests that coping with emergency situations should be narrowed to people who are in the epicenter of happenings. Yet, this attribute may also help to tailor standard mobile crowdsourcing.

# 4  Mobile crowdsourcing model

Because what differs mobile crowdsourcing from the classic models is the availability, reaction time, and mobility, in this paper we introduce geological location as a new important dimension in a crowdsourcing research. Mobile devices have very fast time of HCI and they are not suppressed by 3rd hardware. Analyzing data collected at the server side can significantly improve understanding of a potential hidden behind mobile users.

## 4.1  Work type

In our crowdsourcing model, we distinguish between normal crowdsourcing tasks and emergency situation tasks. Normal tasks e.g. are: translations, describing objects and pictures, citizen science and global problem solving, crowdsourced policy-making and digital deliberation. Emergency situations require immediate action and less thinking, in fact resolving emergency situations happens through crisis mapping and crowdsourced disaster relief. It may be also possible with creating virtual internet of things and creating sensor network ad-hoc. Tasks which are in the middle, are for example: problem of supporting handicapped people. All them open possibility of data gathering by proper organization, and later classifying phone activity and sensory metadata for better situation management in the future events.

## 4.2 Quality model

Hung, Tam et.al (2013) in their paper "*An Evaluation of Aggregation Techniques in Crowdsourcing*" show results from extensible benchmarking framework on the accuracy and robustness to spammers. Scholars explain iterative and non-iterative aggregations of answers. Baba et.al in article "*Statistical quality estimation for general crowdsourcing tasks*" (2013) explain that common approach to tackle the problem of not capable nor motivated workers is to introduce redundancy. The mobile environment will not be free from malicious and inexperienced users, that is why it is necessary to handle biased data. The quality model in our research is a version of model proposed by Hung et.al., together with small modifications. Evaluation measures and crowdsourcing dimensions are explained as follows. Personal response time (PRS) is a time between reading a job description and sending an answer. Hence the PRS attribute can be merged together with the geolocation and computed user profile. Aggregated response time (ARST) is a summary response time per task but from multiple workers. Optional thing to check may be computation time, a simple metric for evaluating aggregation techniques. Obviously, the most important aspect of an aggregation technique is its accuracy. It is straightforward how to measure that—accuracy is defined as the percentage of input objects that are correctly labelled:

**accuracy** = #correctly labelled objects / #total objects

The higher accuracy, the higher power of aggregation method. In experiments, we measure accuracy of each method while varying the number of answers per question and the number of questions per worker. In that, we find which algorithm requires least answers and which algorithm requires least workers to achieve the accuracy requirements. Next attribute is robustness to spammers. In reality, spammers always exist in online community, especially crowdsourcing. Many experiments in the literature showed that the proportion of spammers could be up to 40%, but for the mobile, the ratio is unknown. As a result, it is important for crowdsourcing applications to know how each aggregation technique performs when the worker answers are not trustworthy. Adaptivity to multi-labelling – it is important to know adaptivity of aggregation techniques to multiple-choice questions and multi-answer tasks. Moreover, willingness in choosing multi-label jobs is a good user profiling attribute.

## 4.3 Quality model in dynamic crowds

In paper the "*Quality Control for Real-time Ubiquitous Crowdsourcing*", Mashhadi et.al. (2011) proposed a quality model for dynamic crowds. Researchers showed how to use user mobility pattern to calculate a credibility weight.

## 4.4 Motivation and incentive for workers

Previous studies have identified three broad approaches to motivating contributors: economic incentives, social incentives and intrinsic incentives. In our model we add a gamification model, which rates users based on their efficiency and accuracy, and promotes users proportionally to their involvement in task resolution. Good gamification is giving recognizable profits to top users and prevents stagnation.

## 4.5 Geolocation attribute

Interesting part of the research is to find good and bad locations for assigning crowdsourcing tasks at those locations. We propose a solution of machine learning with semi-supervision. Algorithm goes as follows, iterate: create a set of location classifications {**SL**} (e.g. {*sport object, school, transport*}) and define variants {**SLV**} (e.g. for sport object – *hall, amusement park, public soccer field*). Clusterize the locations to find efficient and non-efficient locations, which prognoses well and bad for a particular type of task (**TT**), which creates a list of pairs { **(SL), (TT)** }.

## 4.6 Response efficiency

Possible way of evaluating personal response time (PRS) is by introducing a "best waiting factor" (β), which is a maximum threshold for which we expect user will give an answer. T means actual waiting time for an answer. The equation for personal response time

**PRS** = $\beta * (1 / t)$

In a case of change in the distance towards β ($\Delta = |\beta - t|$), the evaluation of personal response time (PRS) increases or decreases fractionally.

# 5 Implementation of experiments

There are advantages of bundling the mechanism in OS over allowing an app to listen for requests. A bundled app/module will build a trust from the mobile provider, and will make for easy integration of two parts – crowdsourcing and emergency mode. It is possible to use small network packets to check sensor status of the device, for: better predicting network coverage, for eventually toggling on better devices localization in case of emergency situations, to inform user that he may reach a zone without connection soon (due to analysis of his movement), for statistical analysis of hardware and apps capabilities per region, and finally to enable coping with emergency situations. Example of a good light data propagation is *Amazon Whispernet*.

## 5.1 Result set

There are at least 3 categories, which are aspects for data collection analysis, in a way of supporting this research, but also enabling more further analysis in the future. First is the attribute of geolocation, which we explained more in the paragraph 4.5. Secondly, server side computations can help to choose best aggregations techniques for exactly mobile crowdsourcing, because the dataset will vary from the data collected from e.g. *Amazon Mechanical Turk*. Finally, analyzing this big bunch of data allows to create mobile user profiles, which, when added as an attribute, can help for better task assignment.

# 6 Conclusions

Cyber network of mobile devices users is a promising environment for harvesting knowledge of crowds. It is also a good method for coping with more or less extreme emergency situations, when action is required immediately, and phone users can create ad-hoc a network of self-relief. Despite the controversy of tracking end-users – when overcome – users will provide rich interesting big data from which they will only benefit in the near future.